\documentclass[conference]{IEEEtran}
\newif\ifpdf
\ifx\pdfoutput\undefined
  \pdffalse
\else
  \ifnum\pdfoutput=1
    \pdftrue
  \else
    \pdffalse
  \fi
\fi
\usepackage{amsmath,epsfig, cite, cases}
\usepackage{subfigure,wrapfig}
\usepackage{amssymb,amscd,mathrsfs}
\usepackage[active]{srcltx}
\usepackage{color}
\usepackage{setspace}
\usepackage{enumerate}
\usepackage{bbding}
\usepackage{algorithm, algorithmic}
\usepackage[cm]{fullpage}

\newtheorem{thm}{Theorem}[section]

\newtheorem{lem}[thm]{Lemma}

\newtheorem{define}{Definition}

\def\pc{H}
\def\gen{G}
\def\bh{\zeta}

\def\Fbar{\overline{F}}

\def\E{\mathbb{E}}
\def\mE{\mathcal{E}}

\def\Re{{\mathbb{R}}}
\def\mones{\mathbf{1}_m}

\newcommand{\argmin}{\operatornamewithlimits{argmin}}

\newcommand{\bra}[1]{\left(#1\right)}

\providecommand{\OO}[1]{\operatorname{O}\bigl(#1\bigr)}

\title{Anytime Reliable Codes for Stabilizing Plants over Erasure Channels}

\author{\IEEEauthorblockN{Ravi Teja Sukhavasi} \and \IEEEauthorblockA{Babak Hassibi}
}
\begin{document}

\maketitle
\thispagestyle{empty}
\pagestyle{empty}

\begin{abstract}
The problem of stabilizing an unstable plant over a noisy communication link is an increasingly important one that arises in problems of distributed control and networked control systems. Although the work of Schulman and Sahai over the past two decades, and their development of the notions of \lq\lq tree codes\rq\rq\phantom{} and \lq\lq anytime capacity\rq\rq, provides the theoretical framework for studying such problems, there has been scant practical progress in this area because explicit constructions of tree codes with efficient encoding and decoding did not exist. To stabilize an unstable plant driven by bounded noise over a noisy channel one needs real-time encoding and real-time decoding and a reliability which increases exponentially with delay, which is what tree codes guarantee. We prove the existence of linear tree codes with high probability and, for erasure channels, give an explicit construction with an expected encoding and decoding complexity that is constant per time instant. We give sufficient conditions on the rate and reliability required of the tree codes to stabilize vector plants and argue that they are asymptotically tight. This work takes a major step towards controlling plants over noisy channels, and we demonstrate the efficacy of the method through several examples. 
\end{abstract}

\section{Introduction}
\label{sec: Introduction}
In control theory, the output of a dynamical system is observed and a controller is designed to regulate its behavior. The controller needs to react and generate control signals in real-time. In most traditional control systems, the controller and the plant are colocated and hence there is no measurement loss. There are increasingly many applications such as networked control systems \cite{ncs} and distributed computing \cite{Schulman} where systems are remotely controlled and where measurement and control signals are transmitted across noisy channels. This necessitates a need to \textit{reliably} communicate the measurement and control signals by correcting for the errors introduced by the channels. Although Shannon's information theory is concerned with reliable transmission of a message from one point to another over a noisy channel, the reliability is achieved at the price of large delays which may lead to instability when they occur in the feedback loop of a control system. Hence, we need practical real-time encoding and decoding schemes with appropriate reliability for controlling systems over lossy networks.

Consider a control system with a single observer that communicates with the controller over a lossy communication channel and where the feedback link from the controller to the plant is noiseless. When the channel is rate-limited and deterministic, significant progress has been made (see eg.,\cite{Nair, Matveev}) in understanding the bandwidth requirements for stabilizing open loop unstable systems. When the communication channel is stochastic, \cite{Sahai} provides a necessary and sufficient condition on the communication reliability needed over such a channel to stabilize an unstable scalar linear process, and proposes the notion of feedback anytime capacity as the appropriate figure of merit for such channels. In essence, the encoder is causal and the probability of error in decoding a source symbol that was transmitted $d$ time instants ago should decay exponentially in the decoding delay $d$. 

Although the connection between communication reliability and control is clear, very little is known about error-correcting codes that can achieve such reliabilities. Prior to the work of \cite{Sahai}, and in a different context, \cite{Schulman} proved the existence of codes which under maximum likelihood decoding achieve such reliabilities and referred to them as tree codes. Note that any real-time error correcting code is causal and since it encodes the entire trajectory of a process, it has a natural tree structure to it. \cite{Schulman} proves the existence of nonlinear tree codes yet gives no explicit constructions and/or efficient decoding algorithms. Much more recently \cite{Ostrovsky} proposed efficient error correcting codes for unstable systems where the state grows only polynomially large with time.  So, for linear unstable systems that have an exponential growth rate, all that is known in the way of error correction is the existence of tree codes which are, in general, non-linear. Moreover, the existence results are not with a \lq\lq high probability\rq\rq\phantom{}. When the state of an unstable scalar linear process is available at the encoder, \cite{Yuksel} and \cite{Simsek} develop encoding-decoding schemes that can stabilize such a process over the binary symmetric channel and the binary erasure channel respectively. But little is known in the way of stabilizing partially observed vector-valued processes over a stochastic communication channel.

The subject of error correcting codes for control is in its relative infancy, much as the subject of block coding was after Shannon's seminal work in \cite{Shannon}. So, a first step towards realizing practical encoder-decoder pairs with anytime reliabilities is to explore linear encoding schemes. We consider rate $R=\frac{k}{n}$ causal linear codes which map a sequence of $k$-dimensional binary vectors $\{b_\tau\}_{\tau=0}^\infty$ to a sequence of $n-$dimensional binary vectors $\{c_\tau\}_{\tau=0}^\infty$ where $c_t$ is only a function of $\{b_\tau\}_{\tau=0}^t$. Such a code is anytime reliable if there exist constants $\beta > 0,\eta > 0$ and a delay $d_o > 0$ such that at all times $t$, $P\bigl(\hat{b}_{t-d|t}\neq b_{t-d}\bigr) \leq \eta 2^{-\beta nd}$. 

The contributions of this paper are as follows: 1. We show that linear tree codes exist and further, that they exist with a high probability. 2. For the binary erasure channel, we propose a maximum likelihood decoder whose average complexity of decoding is constant per each time iteration and for which the probability that the complexity at a given time $t$ exceeds $KC^3$ decays exponentially in $C$. 3. We also prove asymptotically tight sufficient conditions on the rate $R$ and exponent $\beta$ needed to stabilize vector-valued processes over a noisy channel. As a consequence, we can efficiently stabilize a partially observed unstable linear process over a binary erasure channel without any channel feedback. 

In Section \ref{sec: problemSetup}, we introduce the notation and set up the problem. In Section \ref{sec: toeplitz}, we introduce the ensemble of time invariant codes and show that they are anytime reliable with a high probability. In Section \ref{sec: DecodingBEC}, we present a simple decoding algorithm for the BEC and in Section \ref{sec: sufficient}, we derive sufficient conditions for stabilizing unstable linear systems over noisy channels. We present some simulations in Section \ref{sec: Simulations} to demonstrate the efficacy of the decoding algorithm.
\section{Problem Setup}
\label{sec: problemSetup}

\begin{figure}
\includegraphics[scale=0.25]{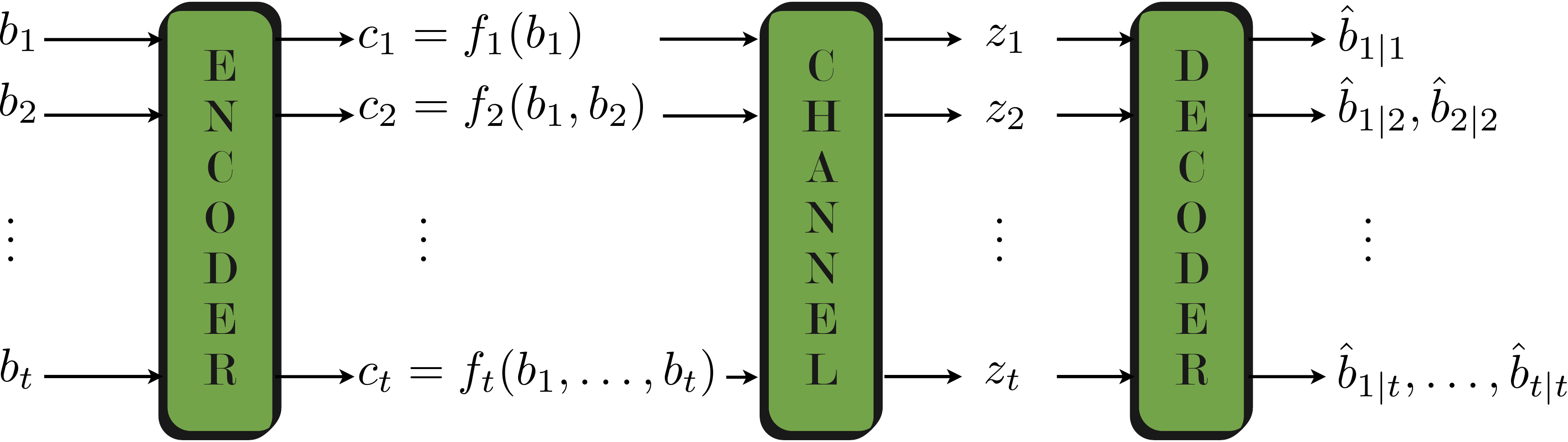}
\caption{Causal encoding and decoding}
\label{fig: operation}
\end{figure} 

We will begin by introducing some notation
\begin{enumerate}
 \item For any matrix $F$, $\Fbar\triangleq \text{abs}(F)$, i.e., $\Fbar_{i,j}=|F_{i,j}|$.$\forall$ $i,j$
 \item $\lambda(F)\triangleq$ largest eigen value of $F$ in magnitude.
 \item For a vector $x$, $x^{(i)}$ denotes the $i^{th}$ component of $x$.
 \item $\mathbf{1}_m \triangleq [1, \ldots, 1]^T$, i.e., a column with $m$ 1's.
 \item For $w,v\in\Re^m$, $w \gtrless v$ denotes component-wise inequality.
\end{enumerate}
Consider the following $m-$dimensional unstable linear system with scalar measurements. Assuming that the system is observable, without loss of generality, it can be cast in the \textit{following} canonical form. 
\begin{align}
\label{eq: sysmodel}
x_{t+1} = Fx_t + Bu_t + w_t,\quad y_t = Hx_t + v_t 
\end{align}
where 
\begin{align*}
 F = \left[\begin{array}{ccccc}
		-a_1 & 1 & 0 & \ldots & \\
		-a_2 & 0 & 1 & 0 &\\
		\vdots & \vdots & &\ddots & \\
		-a_{m-1} & \ldots & \ldots&0 & 1\\
		-a_m & 0 & \ldots & \ldots & 0
		\end{array}\right],\quad H = [1,0,\ldots,0]
\end{align*}
where $\lambda(F)>1$, $u_t$ is the control input and, $w_t$ and $v_t$ are bounded process and measurement noise variables, i.e., $\Vert w_t\Vert_\infty < \frac{W}{2}$ and $\Vert v_t\Vert_\infty < \frac{V}{2}$. Note that the characteristic polynomial of $F$ is $z^n + a_1z^{n-1}+\ldots+a_m$. 

The measurements $\{y_t\}$ are made by an observer while the control inputs $\{u_t\}$ are applied by a remote controller that is connected to the observer by a noisy communication channel. Naturally, the measurements $y_{0:t-1}$ will need to be encoded by the observer to provide protection from the noisy channel while the controller will need to decode the channel outputs to estimate the state $x_t$ and apply a suitable control input $u_t$. This can be accomplished by employing a channel encoder at the observer and a decoder at the controller. For simplicity, we will assume that the channel input alphabet is binary. Suppose one time step of system evolution in \eqref{eq: sysmodel} corresponds to $n$ channel uses\footnote{In practice, the system evolution in \eqref{eq: sysmodel} is obtained by discretizing a continuous time differential equation. So, the interval of discretization could be adjusted to correspond to an integer number of channel uses, provided the channel use instances are close enough.}. Then, at each instant of time $t$,  the operations performed by the observer, the channel encoder,  the channel decoder and the controller can be described as follows. The observer generates a $k-$bit message, $b_t\in\{0,1\}^k$, that is a causal function of the measurements, i.e., it depends only on $y_{0:t}$. Then the channel encoder causally encodes $b_{0:t} \in \{0,1\}^{kt}$ to generate the $n$ channel inputs $c_t\in\{0,1\}^n$. Note that the rate of the channel encoder is $R = k/n$. Denote the $n$ channel outputs corresponding to $c_t$ by $z_t \in \mathcal{Z}^n$, where $\mathcal{Z}$ denotes the channel output alphabet. Using the channel outputs received so far, i.e., $z_{0:t}\in \mathcal{Z}^{nt}$, the channel decoder generates estimates $\{\hat{b}_{\tau|t}\}_{\tau \leq t}$ of $\{b_\tau\}_{\tau\leq t}$, which, in turn, the controller uses to generate the control input $u_{t+1}$. This is illustrated in Fig. \ref{fig: operation}. \textit{Note that we do not assume any channel feedback}. Now, define
\begin{align*}
 P_{t,d}^e = P\bra{\min\{\tau:\hat{b}_{\tau|t}\neq b_\tau\} = t-d+1}
\end{align*}
Thus, $P^e_{t,d}$ is the probability that the earliest error is $d$ steps in the past.
\begin{define}[Anytime reliability]
 We say that an encoder-decoder pair is $(R,\beta,d_o)-$anytime reliable if
\begin{align}
\label{eq: anytimeReliability}
 P_{t,d}^e \leq \eta 2^{-n\beta d},\,\,\,\forall\,\,\,t,d\geq d_o
\end{align}
In some cases, we write that a code is $(R,\beta)-$anytime reliable. This means that there exists a fixed $d_o > 0$ such that the code is $(R,\beta,d_o)-$anytime reliable.
\end{define}

We will show in Section \ref{sec: sufficient} (Theorem \ref{thm: cuboidalThm}) that $(R,\beta)-$anytime reliability is a sufficient condition to stabilize \eqref{eq: sysmodel} in the mean squared sense\footnote{can be easily extended to any other norm}. In what follows, we will demonstrate causal linear codes which under maximum likelihood decoding achieve such exponential reliabilities.
\section{Linear Anytime Codes}
\label{sec: toeplitz}
As discussed earlier, a first step towards developing practical encoding and decoding schemes for automatic control is to study the existence of linear codes with anytime reliability. We will begin by defining a causal linear code.
\begin{define}[Causal Linear Code]
 A causal linear code is a sequence of linear maps $f_\tau:\{0,1\}^{k\tau}\mapsto \{0,1\}^n$, $\tau\geq 0$ and hence can be represented as
\begin{align}
\label{eq: GnR}
 f_\tau(b_{1:\tau}) = \gen_{\tau 1}b_1 + \gen_{\tau 2}b_2 + \ldots + \gen_{\tau \tau}b_\tau
\end{align}
where $\gen_{ij}\in\{0,1\}^{n\times k}$
\end{define}
We denote $c_\tau \triangleq f_\tau(b_{1:\tau})$. Note that a tree code is a more general construction where $f_\tau$ need not be linear. Also note that the associated code rate is $R = \frac{k}{n}$.  The above encoding is equivalent to using a semi-infinite dimensional block lower triangular generator matrix, $G_{n,R}$, whose entries are clear from \eqref{eq: GnR}
or equivalently as a semi-infinite dimensional block lower triangular parity check matrix, $\mathbb{H}_{n,R}$ (the parity check matrix satisfies $H_{n,R}G_{n,R}=0$.)
\begin{align}
\label{eq: paritycheck}
  \mathbb{H}_{n,R} = \left[\begin{array}{ccccc}
				\pc_{11} & 0 & \ldots & \ldots & \ldots \\
				\pc_{21} & \pc_{22} & 0 & \ldots & \ldots \\
				\vdots & \vdots & \ddots & \vdots & \vdots \\
				\pc_{\tau 1} & \pc_{\tau 2} & \ldots & \pc_{\tau\tau} & 0\\
				\vdots & \vdots & \vdots & \vdots & \ddots
				\end{array}\right]
\end{align}
where\footnote{While for a given generator matrix, the parity check matrix is not unique, when $\mathbb{G}_{n,R}$ is block lower, it is easy to see that $\mathbb{H}_{n,R}$ can also be chosen to be block lower.} $\pc_{ij}\in\{0,1\}^{\overline{n}\times n}$ and $\overline{n}=n(1-R)$. In order to ensure that the code rate is equal to the design rate $R = \frac{k}{n}$, $\mathbb{H}^t_{n,R}$ needs to be full rank for every $t$, where $\mathbb{H}^t_{n,R}$ is the $\overline{n}t\times nt$ leading principal minor of $\mathbb{H}_{n,R}$. This will happen if $H_{ii}$ is full rank for all $i$. The existence results that follow imply the existence of anytime reliable $\mathbb{H}_{n,R}$ whose code rate is same as the design rate.

We will present all our results for binary input output symmetric channels\footnote{which can be easily extended to more general memoryless channels}. The Bhattacharya parameter $\bh$ for such channels is defined as
\begin{align*}
 \bh = \int\limits_{-\infty}^\infty\sqrt{p\bra{z|X=1}p\bra{z|X=0}}dz
\end{align*}
where $z$ and $X$ denote the channel output and input, respectively. We will begin by proving the existence of such codes that are $(R,\beta)-$anytime reliable over a finite time horizon, $T$, i.e., under ML decoding $P_{d,t}^e \leq \eta 2^{-\beta d},\,\,\,\forall \,\,\,t\leq T$. We will then prove their existence for all time. Due to space limitations, proofs for all the results in this section are presented in a companion paper, \cite{ISIT}.

\subsection{Finite Time Horizon}
Over a finite time horizon, $T$, a causal linear code is represented by a block lower triangular parity check matrix $\mathbb{H}_{n,R,T}\in\{0,1\}^{\overline{n}T\times nT}$. The following Theorem guarantees the existence of a $\mathbb{H}_{n,R,T}$
 that is $(R,\beta)-$anytime relable.
\begin{thm}
\label{thm: finiteTimeHorizon}
For each time $T > 0$, rate $R$ and exponent $\beta$ such that
\begin{align*}
 R &< 1- \log_{2}(1+\bh),\quad\text{and}\\
\beta  &< H^{-1}(1-R)\bra{\log_2\bra{\frac{1}{\bh}} + \log_2\bigl[2^{1-R}-1\bigr]}
\end{align*}
there exists a causal linear code $H(n,k,T)$ that is $(R,\beta)-$anytime reliable.
\end{thm}
$H^{-1}(1-R)$ is the smaller root of the equation $H(x)=1-R$, where $H(.)$ is the binary entropy function. Theorem \ref{thm: finiteTimeHorizon} proves the existence of finite dimensional causal linear codes, $\mathbb{H}_{n,R,T}$, that are anytime reliable for decoding instants upto time $T$. In the following subsection, we demonstrate the existence of semi-infinite causal linear codes, $\mathbb{H}_{n,R}$, that are anytime reliable for all decoding instants. We also show that such codes drawn from an appropriate ensemble are anytime reliable with a high probability. The key is to impose a Toeplitz structure on the parity check matrix. 

\subsection{Time Invariant Codes}
Consider causal linear codes with the following Toeplitz structure
\begin{align*}
   \mathbb{H}_{n,R}^{TZ} = \left[\begin{array}{ccccc}
				\pc_{1} & 0 & \ldots & \ldots & \ldots \\
				\pc_{2} & \pc_{1} & 0 & \ldots & \ldots \\
				\vdots & \vdots & \ddots & \vdots & \vdots \\
				\pc_{\tau} & \pc_{\tau-1} & \ldots & \pc_{1} & 0\\
				\vdots & \vdots & \vdots & \vdots & \ddots
				\end{array}\right]
\end{align*}
The superscript $TZ$ in $\mathbb{H}_{n,R}^{TZ}$ denotes \lq Toeplitz'. $\mathbb{H}_{n,R}^{TZ}$ is obtained from $\mathbb{H}_{n,R}$ in \eqref{eq: paritycheck} by setting $\pc_{ij} = \pc_{i-j+1}$ for $i\geq j$. Due to the Toeplitz structure, we have the following invariance, $P_{t,d}^e = P_{t',d}^e$ for all $t,t'$. The notion of time invariance is analogous to the convolutional structure used to show the existence of infinite tree codes in \cite{Schulman}. The code $\mathbb{H}_{n,R}^{TZ}$ will be referred to as a time-invariant code. This time invariance obviates the need to union bound over all time $t$ and hence allows us to prove that such codes which are anytime reliable are abundant. 

\begin{define}[The ensemble $\mathbb{TZ}_p$]
 The ensemble $\mathbb{TZ}_p$ of time-invariant codes, $\mathbb{H}_{n,R}^{TZ}$, is obtained as follows, $\pc_1$ is any full rank binary matrix and for $\tau \geq 2$, the entries of $H_\tau$ are chosen i.i.d according to Bernoulli($p$), i.e., each entry is 1 with probability $p$ and 0 otherwise.
\end{define}
Note that $H_1$ being full rank implies that $H^t_{n,R}$ is full rank for every $t$. For the ensemble $\mathbb{TZ}_p$, we have the following result

\begin{thm}[Abundance of time-invariant codes]
 \label{thm: Toeplitz}
For any rate $R$ and exponent $\beta$ such that
\begin{subequations}
\label{eq: thresholdsBEC}
\begin{align*}
 R &< 1- \frac{\log_{2}(1+\bh)}{\log_2(1/(1-p))},\quad\text{and}\\
\beta  &< H^{-1}(1-R)\bra{\log_2\bra{\frac{1}{\bh}} + \log_2\bigl[(1-p)^{-(1-R)}-1\bigr]}
\end{align*}
\end{subequations}
if $\mathbb{H}_{n,R}^{TZ}$ is chosen from $\mathbb{TZ}_p$, then
\begin{align*}
 P\bra{\mathbb{H}_{n,R}^{TZ}\text{ is }(R,\beta,d_o)-\text{anytime reliable}} \geq 1-2^{-\Omega(nd_o)}
\end{align*}
\end{thm}

Note that by choosing $p$ small, we can trade off better rates and exponents with sparser parity check matrices. Note that for BEC($\epsilon$), $\bh = \epsilon$ and for BSC($\epsilon$), $\bh = 2\sqrt{\epsilon(1-\epsilon)}$. For the Binary Symmetric Channel (BSC) with bit flip probability $\epsilon$ and for $p=\frac{1}{2}$, the threshold for rate in Theorem \ref{thm: Toeplitz} becomes $R < 1-2\log_2(\sqrt{\epsilon}+\sqrt{1-\epsilon})$. It turns out that this can be strengthened as follows.
\begin{thm}[Tighter bounds for BSC($\epsilon$)]
\label{thm: tighterBSC}
For any rate $R$ and exponent $\beta$ such that
\begin{align*}
 R < 1 - H(\epsilon),\,\,\,\beta  < KL\bra{H^{-1}(1-R)\| \min\{\epsilon,1-\epsilon\}}
\end{align*}
if $\mathbb{H}_{n,R}^{TZ}$ is chosen from $\mathbb{TZ}_{\frac{1}{2}}$, then
\begin{align*}
 P\bra{\mathbb{H}_{n,R}^{TZ}\text{ is }(R,\beta,d_o)-\text{anytime reliable}} \geq 1-2^{-\Omega(nd_o)}
\end{align*}
\end{thm}

\section{Decoding over the BEC}
\label{sec: DecodingBEC}

Owing to the simplicity of the erasure channel, it is possible to come up with an efficient way to perform maximum likelihood decoding at each time step. We will show that the average complexity of the decoding operation at any time $t$ is constant and that it being larger than $KC^3$ decays exponentially in $C$. Consider an arbitrary decoding instant $t$, let $c=[c_1^T,\ldots,c_t^T]^T$ be the transmitted codeword and let $z=[z_1^T,\ldots,z_t^T]^T$ denote the corresponding channel outputs. Recall that $\mathbb{H}_{n,R}^t$ denotes the $\overline{n}t\times nt$ leading principal minor of $\mathbb{H}_{n,R}$. Let $z_e$ denote the erasures in $z$ and let $H_e$ denote the columns of $\mathbb{H}_{n,R}^t$  that correspond to the positions of the erasures. Also, let $\tilde{z}_e$ denote the unerased entries of $z$ and let $\tilde{H}_e$ denote the columns of $\mathbb{H}_{n,R}^t$ excluding $H_e$. So, we have the following parity check condition on $z_e$, $H_ez_e = \tilde{H}_e\tilde{z}_e$. Since $\tilde{z}_e$ is known at the decoder, $s\triangleq \tilde{H}_e\tilde{z}_e$ is known. Maximum likelihood decoding boils down to solving the linear equation $H_ez_e = s$.
Due to the lower triangular nature of $H_e$, unlike in the case of traditional block coding, this equation will typically not have a unique solution, since $H_e$ will typically not be full rank. This is alright as we are not interested in decoding the entire $z_e$ correctly, we only care about decoding the earlier entries accurately. If $z_e = [z_{e,1}^T,\,\, z_{e,2}^T]^T$, then $z_{e,1}$ corresponds to the earlier time instants while $z_{e,2}$ corresponds to the latter time instants. The desired reliability requires one to recover $z_{e,1}$ with an exponentially smaller error probability than $z_{e,2}$. Since $H_e$ is lower triangular, we can write $H_ez_e = s$ as
\begin{align}
\label{eq: bec1}
 \left[\begin{array}{cc}
  H_{e,11} & 0\\
  H_{e,21} & H_{e,22}
 \end{array}\right]\left[\begin{array}{c}z_{e,1}\\z_{e,2}\end{array}\right] = \left[\begin{array}{c}s_1\\s_2\end{array}\right]
\end{align}
Let $H_{e,22}^\bot$ denote the orthogonal complement of $H_{e,22}$, ie., $H_{e,22}^\bot H_{e,22} = 0$. Then multiplying both sides of \eqref{eq: bec1} with diag$(I,H_{e,22})$, we get
\begin{align}
 \label{eq: bec2}
 \left[\begin{array}{c}
  H_{e,11}\\
  H_{e,22}^\bot H_{e,21}
 \end{array}\right]z_{e,1} = \left[\begin{array}{c}s_1\\H_{e,22}^\bot s_2\end{array}\right]
\end{align}
If $[H_{e,11}^T\,\,\, (H_{e,22}^\bot H_{e,21})^T]^T$ has full column rank, then $z_{e,1}$ can be recovered exactly. The decoding algorithm now suggests itself, i.e., find the smallest possible $H_{e,22}$ such that $[H_{e,11}^T\,\,\, (H_{e,22}^\bot H_{e,21})^T]^T$ has full rank and it is outlined in Algorithm \ref{alg: algorithm}.
\begin{algorithm}
\caption{Decoder for the BEC}
\label{alg: algorithm}
\begin{enumerate}
\item Suppose, at time $t$, the earliest uncorrected error is at a delay $d$. Identify $z_e$ and $H_e$ as defined above.
\item Starting with $d'=1,2,\ldots,d$, partition
\begin{align*}
 z_e = [z_{e,1}^T\,\,z_{e,2}^T]^T\,\,\text{and}\,\,H_e = \left[\begin{array}{cc}H_{e,11}&0\\H_{e,21}&H_{e,22}\end{array}\right]
\end{align*}
where $z_{e,2}$ correspond to the erased positions up to delay $d'$.
\item Check whether the matrix $\left[\begin{array}{c}
  H_{e,11}\\
  H_{e,22}^\bot H_{e,21}
 \end{array}\right]$
has full column rank.
\item If so, solve for $z_{e,1}$ in the system of equations
\begin{align*}
  \left[\begin{array}{c}
  H_{e,11}\\
  H_{e,22}^\bot H_{e,21}
 \end{array}\right]z_{e,1} = \left[\begin{array}{c}s_1\\H_{e,22}^\bot s_2\end{array}\right]
\end{align*}
\item Increment $t=t+1$ and continue.
\end{enumerate}
\end{algorithm} 

\subsection{Complexity}
Suppose the earliest uncorrected error is at time $t-d+1$, then steps 2), 3) and 4) in Algorithm \ref{alg: algorithm} can be accomplished by just reducing $H_e$ into the appropriate row echelon form, which has complexity $\OO{d^3}$. The earliest entry in $z_e$ is at time $t-d+1$ implies that it was not corrected at time $t-1$, the probability of which is $P_{d-1,t-1}^e \leq \eta 2^{-n\beta (d-1)}$. Hence, the average decoding complexity is at most $K\sum_{d>0}d^3 2^{-n\beta d}$ which is bounded and is independent of $t$. In particular, the probability of the decoding complexity being $Kd^3$ is at most $\eta 2^{-n\beta d}$. The decoder is easy to implement and its performance is simulated in Section \ref{sec: Simulations}. Note that the encoding complexity per time iteration increases linearly with time. This can also be made constant on average if the decoder can send periodic acks back to the encoder with the time index of the last correctly decoded source bit. 
\section{Sufficient Conditions for Stabilizability}
\label{sec: sufficient}

Consider an unstable $m-$dimensional linear system whose state space equations in canonical form are given by \eqref{eq: sysmodel}, i.e., $\lambda(F) > 1$, and recall that the characteristic polynomial of $F$ is $z^n + a_1z^{n-1}+\ldots+a_m$.  Suppose the observer does not have any feedback from the controller, in particular, it does not have access to the control inputs. Then we can stabilize such a system in the mean squared sense over a noisy channel provided that the rate $R$ and exponent $\beta$ of the $(R,\beta)-$anytime reliable code used to encode the measurements satisfy the following sufficient condition.

\begin{thm}[No Feedback to the Observer]
\label{thm: cuboidalThm}
 It is possible to stabilize \eqref{eq: sysmodel} in the mean squared sense with an $(R,\beta)-$anytime code provided $(F,B)$ is controllable and 
\begin{align}
\label{eq: cuboidalThm}
 R > R_{n} = \frac{1}{n}\log_2\sum_{i=1}^m |a_i|,\,\,\,\,\beta > \beta_{n} = \frac{2}{n}\log_2\lambda(\Fbar)
\end{align}
\end{thm}

If the observer knows the control inputs, it turns out that one can make do with lower rates. This is stated as the following Theorem

\begin{thm}[Observer Knows the Control Inputs]
\label{thm: cuboidalThmWF}
 When the observer has access to the control inputs, it is possible to stabilize \eqref{eq: sysmodel} in the mean squared sense with an $(R,\beta)-$anytime code provided $(F,B)$ is controllable and 
\begin{subequations}
\label{eq: cuboidalThmWF}
\begin{align}
 R &> R^f_{n} = \argmin_{r}\left\{\lambda(\Fbar D_{nr}) < 1\right\}\\
 \beta &> \beta^f_{n} = \frac{2}{n}\log_2\lambda(\Fbar)
\end{align}
\end{subequations}
where $D_{nr} = \text{diag}\bra{2^{-nr},1,\ldots,1}$. Moreover
\begin{align}
\label{eq: ubRfcn}  R^f_{n} \leq \frac{1}{n}\log_2\max\left\{|a_m|2^{m-1},\max_{1\leq i\leq m-1}|a_i|2^i\right\}
\end{align}
\end{thm}

The superscript $f$ in $R^f_{n}$ denotes \lq feedback' to emphasize the fact that the observer has access to the control inputs. 
Before proceeding further, we will give a brief outline of the proofs for Theorems \ref{thm: cuboidalThm} and \ref{thm: cuboidalThmWF} (details are in Section \ref{sec: proofCuboidal}). At each time $t$, using the channel outputs received received till $t$, we bound the set of all possible states that are consistent with the estimates of the quantized measurements using a hypercuboid, i.e., a region of the form $\left\{x_t\in\Re^m | x_{min,t|t}\leq x_t\leq x_{max,t|t}\right\}$, where $x_{min,t|t},x_{max,t|t}\in\Re^m$ and the inequalities are component-wise. If $\Delta_{t|t} = x_{max,t|t}-x_{min,t|t}$, then from Lemma \ref{lem: cubTimeUpdate}, $\Delta_{t+1|t} = \Fbar\Delta_{t|t} + W\mones$. The anytime exponent is determined by the growth of $\Delta_t$ in the absence of measurements, hence the bound $\beta_{n} = \beta^f_{n} = 2\log_2\lambda(\Fbar)$. The bound on the rate is determined by how fine the quantization needs to be for $\Delta_t$ to be bounded asymptotically. 

\subsection{The Limiting Case}

The sufficient conditions derived above are for the case when the measurements are encoded every time step. Alternately, one can encode the measurements every, say $\ell$, time steps, and consider the asymptotic rate and exponent needed as $\ell$ grows. Note that this amounts to working with the system matrix $F^\ell$. So, one can calculate this limiting rate and exponent by writing the eigen values of $F$, $\{\lambda_i\}_{i=1}^m$, as $\lambda_i = \mu_i^n$ and letting $n$ scale. The following asymptotic result allows us to compare the sufficient conditions above with those in the literature (eg., see \cite{Sahai,Nair,Minero}). 
\begin{thm}[The Limiting Case]
\label{thm: limitingCase}
Write the eigen values of $F$, $\{\lambda_i\}_{i=1}^m$, in the form $\lambda_i = \mu_i^{n}$. Letting $n$ scale, $R_{n}$ and $R^f_{n}$ converge to $R^*$, and $\beta_{n}$ and $\beta^f_{n}$ converge to $\beta^*$, where
\begin{align}
\label{eq: limiting} R^* =\sum_{i:|\mu_i|>1}\log_2|\mu_i|,\,\,\,\beta^* = 2\log_2\max_i|\mu_i|
\end{align}
In addition, the upper bounds on $R^f_{n}$ in \eqref{eq: ubRfcn} also converges to $R^*$.
\end{thm}
\begin{IEEEproof}
 See Section \ref{subsec: limitingCase} of the Appendix.
\end{IEEEproof}
For stabilizing plants over deterministic rate limited channels, \cite{Nair} showed that a rate $R > R^*$, where $R^*$ is as in \eqref{eq: limiting}, is necessary and sufficient. So, asymptotically the sufficient conditions for the rate $R$ in Theorems \ref{thm: cuboidalThm} and \ref{thm: cuboidalThmWF} are tight. Though the above limiting case allows one to obtain a tight and an intuitively pleasing characterization of the rate and exponent needed, it should be noted that this may not be operationally practical. For, if one encodes the measurements every $\ell$ time steps, even though Theorem \ref{thm: limitingCase} guarantees stability, the performance of the closed loop system (the LQR cost, say) may be unacceptably large because of the delay we incur. This is what motivated us to present the sufficient conditions in the form that we did above. 

\subsection{A Comment on the Trade-off Between Rate and Exponent}

Once a set of rate-exponent pairs $(R,\beta)$ that can stabilize a plant is available, one would want to identify the pair that optimizes a given cost function. Higher rates provide finer resolution of the measurements while larger exponents ensure that the controller's estimate of the plant does not drift away; however, we cannot have both. One can either coarsely quantize the measurements and protect the bits heavily or quantize them moderately finely and not protect the bits as much. One can easily cook up examples using an LQR cost function with the balance going either way. Studying this trade-off is integral to making the results practically applicable.

\section{Tighter Bounds on the Anytime Exponent}
\label{sec: Ellipsoidal}
From Theorem \ref{thm: cuboidalThm}, using the technique outlined in the previous section, one needs an exponent $n\beta \geq 2\log\lambda(\Fbar)$. It turns out that a smaller exponent of $2\log_2\lambda(F)$ suffices. The idea is to alternately bound the set of all possible states that are consistent with the estimates of the quantized measurements using an ellipsoid $\mE(P,c)\triangleq \left\{x\in\Re^m| \langle x-c,P^{-1}(x-c)\rangle \leq 1\right\}$. This can be seen as an extension of the technique proposed in \cite{Schweppe} to filtering using quantized measurements. If $m=1$, $\lambda(\Fbar)=\lambda(F)$. So, let $m\geq 2$. 

In view of the duality between estimation and control, we can focus on the problem of tracking \eqref{eq: sysmodel} over a noisy communication channel. For, if \eqref{eq: sysmodel} can be tracked with an asymptotically finite mean squared error and if $(F,B)$ is stabilizable, then it is a simple exercise to see that there exists a control law $\{u_t\}$ that will stabilize the plant in the mean squared sense, i.e., $\limsup_t\E\Vert x_t\Vert^2 < \infty$. In particular, if the control gain $K$ is chosen such that $\sqrt{2}F + BK$ is stable, then $u_t = K\hat{x}_{t|t}$ will stabilize the plant, where $\hat{x}_{t|t}$ is the estimate of $x_t$ using channel outputs up to time $t$. Hence, in the rest of the analysis, we will focus on tracking \eqref{eq: sysmodel}. The control input $u_t$ therefore is assumed to be absent, i.e., $u_t=0$.

We will first present a recursive state estimation algorithm using the channel outputs and then state the sufficient conditions needed for the estimation error to be appropriately bounded using such a filter. Recall that the channel outputs corresponding to the coded bits $c_t\in GF_2^n$ are $z_t\in\mathcal{Z}^n$. Let $x_0\in\mE(P_0,0)$ and suppose using $\{z_\tau\}_{\tau\leq t-1}$, we have $x_t\in\mE(P_{t|t-1},\hat{x}_{t|t-1})$.  Note that, since $H = [1, 0,\ldots, 0]$, the measurement update provides information of the form $x_{min,t|t}^{(1)}\leq x_t^{(1)}\leq x_{max,t|t}^{(1)}$, which one may call a slab. $\mE(P_{t|t},\hat{x}_{t|t})$ would then be an ellipsoid that contains the intersection of the above slab with $\mE(P_{t|t-1},\hat{x}_{t|t-1})$, in particular one can set it to be the minimum volume ellipsoid covering this intersection. Lemma \ref{lem: minVol} gives a formula for the minimum volume ellipsoid covering the intersection of an ellipsoid and a slab. Note that the width of the slab above tends to be smaller if the observer has access to the control inputs than when it does not. For the time update, it is easy to see that for any $\epsilon > 0$ and $P_{t+1} = (1+\epsilon)FP_{t|t}F^T + \frac{W^2}{4\epsilon}\mones$, $\mE(P_{t+1},F\hat{x}_{t|t})$ contains the state $x_{t+1}$ whenever $\mE(P_{t|t},\hat{x}_{t|t})$ contains $x_t$. This leads to the following Lemma. For convenience, we write $P_{t}$ for $P_{t|t-1}$.
\begin{lem}[The Ellipsoidal Filter]
Whenever $\mE(P_0,0)$ contains $x_0$, for each $\epsilon > 0$, the following filtering equations give a sequence of ellipsoids $\left\{\mE(P_{t|t},\hat{x}_{t|t})\right\}$ that, at each time $t$, contain $x_t$. 
\begin{subequations}
 \begin{align}
  P_{t+1} &= (1+\epsilon)FP_{t|t}F^T + \frac{W^2}{4\epsilon}\mones,\,\,\hat{x}_{t+1} = F\hat{x}_{t|t}\\
  P_{t|t} &= b_tP_t - (b_t-a_t)\frac{P_te_1e_1^TP_t}{e_1^TP_te_1},\,\,\hat{x}_{t|t} = \xi_t\frac{P_te_1}{\sqrt{e_1^TP_te_1}}
 \end{align}
\end{subequations}
where $a_t,b_t$ and $\xi_t$ can be calculated in closed form using Lemma \ref{lem: minVol}.
\end{lem}

Using this approach, we get the following set of sufficient conditions. The proofs are similar to the proofs of Theorems \ref{thm: cuboidalThm} and \ref{thm: cuboidalThmWF}, and hence skipped due to space limitations.
\begin{thm}[No Feedback to the Observer]
\label{thm: ellipsoidalThm}
 It is possible to stabilize \eqref{eq: sysmodel} for $m\geq 2$ in the mean squared sense with an $(R,\beta)-$anytime code provided $(F,B)$ is controllable and 
\begin{subequations}
\label{eq: ellipsoidalThm}
\begin{align}
 R &> R_{e,n} = \frac{1}{n}\log_2\left[\frac{\sqrt{m}}{2}\sum_{i=1}^m |a_i|\theta^{i-1}\right]\\
 \beta &> \beta_{e,n} = \frac{2}{n}\log_2\lambda(F)
\end{align}
\end{subequations}
 where $\theta = \frac{m}{m-1}$
\end{thm}

\begin{thm}[Observer Knows the Control Inputs]
\label{thm: ellipsoidalThmWF}
 When the observer has access to the control inputs, it is possible to stabilize \eqref{eq: sysmodel} in the mean squared sense with an $(R,\beta)-$anytime code provided $(F,B)$ is controllable and 
\begin{subequations}
\label{eq: ellipsoidalThmWF}
\begin{align}
 R &> R^f_{e,n} = \argmin_r \left\{\lambda(\Fbar D_{m,nr})<1\right\}\\
 \beta &> \beta^f_{e,n} = \frac{2}{n}\log_2\lambda(F)
\end{align}
\end{subequations}
 where $D_{m,nr} = \text{diag}\bra{\sqrt{m}2^{-nr},\sqrt{\theta},\ldots,\sqrt{\theta}},\,\,\theta=\frac{m}{m-1}$. Moreover
\begin{align}
\label{eq: ubRfen} R^f_{e,n} &\leq \frac{1}{2n}\log_2m + \nonumber\\ 
 &\frac{1}{n}\log_2\max\left\{|a_m|(2\theta)^{m-1},\max_{1\leq i\leq m-1}2|a_i|(2\theta)^{i-1}\right\}
\end{align}
\end{thm}

In the same limiting sense as described in Section \ref{sec: sufficient}, $R^f_{e,n}$ and $R_{e,n}$ converge to $R^*$ while $\beta^f_{e,n}$ and $\beta_{e,n}$ converge to $\beta^*$, where $R^*$ and $\beta^*$ are as in the Lemma \ref{thm: limitingCase}. The proof is in Section \ref{subsec: limitingCase} of the Appendix.
\section{Proofs of Theorems \ref{thm: cuboidalThm} and \ref{thm: cuboidalThmWF}}
\label{sec: proofCuboidal}

The analysis will proceed in two steps. We will first determine a sufficient condition on the number of bits per measurement, $nR$, that are required to track \eqref{eq: sysmodel} when these bits are available error free. We will then determine the anytime exponent $n\beta$ needed in decoding these source bits when they are communicated over a noisy channel. 

At each time, we bound the set of all possible states that are consistent with the quantized measurements using a hypercuboid, i.e., a region of the form $\left\{x\in\Re^m | x_{min}\leq x\leq x_{max}\right\}$, where $x_{min},x_{max}\in\Re^m$ and the inequalities are component-wise. In what follows, we call $\Delta_{t|\tau}\triangleq x_{max,t|\tau}-x_{min,t|\tau}$, the uncertainty in $x_t$ using $\{b_\tau'\}_{\tau'\leq \tau}$, i.e., quantized measurements up to time $\tau$. For convenience, let $\Delta_t \equiv \Delta_{t|t-1}$. Then, the time update is given by the following Lemma.
\begin{lem}[Time Update]
\label{lem: cubTimeUpdate}
The time update relating $\Delta_{t+1}$ and $\Delta_{t|t}$ is given by $\Delta_{t+1} = \Fbar\Delta_{t|t} + W\mones$
\end{lem}
\begin{IEEEproof}
 From the system dynamics in \eqref{eq: sysmodel}, the following is immediate
\begin{align*}
 \Delta_{t+1}^{(i)} &= W + \max\left\{\left|\pm a_i\Delta_{t|t}^{(1)} + \Delta_{t|t}^{(i+1)}\right|,\left|\Delta_{t|t}^{(i+1)}\right|,\left|a_i\Delta_{t|t}^{(1)}\right|\right\}\\
 &= |a_i|\Delta_{t|t}^{(1)} + \Delta_{t|t}^{(i+1)} + W,\,\,\,i\leq m-1\\
\Delta_{t+1}^{(m)} &= |a_m|\Delta_{t|t}^{(1)} + W
\end{align*}
In short, the above equations amount to $\Delta_{t+1} = \Fbar\Delta_{t|t} + W\mones$.
\end{IEEEproof}

The measurement update depends on whether or not the observer has access to the control inputs. 

\subsection{Observer does not know the control inputs}
In this case, the observer simply quantizes the measurements $y_t$ according to a $2^{nR}-$regular lattice quantizer with bin width $\delta$, i.e., the quantizer is defined by $Q:\Re\mapsto\{0,1,\ldots,2^{nR}-1\}$, where $Q(x) = \lfloor\frac{x}{\delta}\rfloor\text{ mod }2^{nR}$. Assuming that the rate, $R$, is large enough, we will first find the steady state value of the recursion for $\Delta_t$, which we then use to determine $R$. At each time $t$, the observer can communicate the measurement $y_t$ to within an uncertainty of $\delta$, i.e., the estimator knows that the measurement lies in an interval of width $\delta$. Adding to this the effect of the observation noise, $-\frac{V}{2} \leq v_t \leq \frac{V}{2}$, the estimator knows $x_t^{(1)}$ to within an uncertainty of $\Delta_{t|t}^{(1)} = \delta + V$. Note that $\Delta_{t|t}^{(i)} = \Delta_{t}^{(i)}$ for $i\neq 1$. Combining this observation with Lemma \ref{lem: cubTimeUpdate}, the following is fairly straightforward.
\begin{lem}[Steady State value of $\Delta_t$ without feedback]
If $\lim_{t\rightarrow\infty}\Delta_t = \Delta_\infty$, then $\Delta_\infty = (\delta+V)L_ua + WL_u\mones$, where $a = [|a_1|,\ldots,|a_m|]^T$ and $L_u = [\ell_{ij}]_{1\leq i,j\leq m}$ with $\ell_{ij} = \mathbb{I}_{i\leq j}$.
\end{lem}

Now, we need to go back and calculate $R$. Observe that $\Delta_\infty$ does not depend on the starting value $\Delta_0$. So we just need $\delta2^{nR}\geq \max\left\{\Delta_\infty^{(1)},\Delta_0^{(1)}\right\} + V$. From the above Lemma, $\Delta_\infty^{(1)} = \delta\sum_{i=1}^m|a_i| + V\sum_{i=1}^m|a_i|+mW$. So, we need
\begin{align*}
 2^{nR} > \max\left\{\sum_{i=1}^m|a_i| + \frac{V + V\sum_{i=1}^m|a_i| + mW}{\delta}, \frac{\Delta_0^{(1)}}{\delta}\right\}
\end{align*}

The minimum required rate is obtained by letting $\delta\rightarrow\infty$, in which case we need $R > \frac{1}{n}\log_2\sum_{i=1}^m|a_i|$ and this gives $R_{n}$ in Theorem \ref{thm: cuboidalThm}. 

\subsection{Observer knows the control inputs}

In this case, the observer can infer that the uncertainty in $y_t$ at the estimator side is $\Delta_t^{(1)} + V$. So, it can use the $nR$ bits to shrink this to $2^{-nR}(\Delta_t^{(1)}+V)$. Taking into account the observation noise, the uncertainty in $x_t$ after the measurement update will be given by $\Delta_{t|t}^{(1)} = V + 2^{-nR}(\Delta_t^{(1)}+V)$ and $\Delta_{t|t}^{(i)} = \Delta_{t}^{(i)}$ for $i\neq 1$. Combining this with Lemma \ref{lem: cubTimeUpdate}, the overall recursion for $\Delta_t$ is given by
\begin{align}
\label{eq: recurDelta1}
 \Delta_{t+1} &= \Fbar D_{nR}\Delta_t + W_{c,nR},\,\,\,\text{where}\nonumber\\
D_{nR} &= \text{diag}\{2^{-nR},1,\ldots,1\}\nonumber\\
W_{c,nR} &= [V(1+2^{-nR})+W,W,\ldots,W]^T
\end{align}
Noting that $V(1+2^{-nR})\leq 2V$, the above recursion is bounded if and only if $\Fbar D_{nR}$ is stable. It follows that $\Fbar D_{nR}$ is stable for all $R > R^f_{n}$, where $R^f_{n} = \frac{1}{n}\argmin_r\left\{\lambda\bra{\Fbar D_{nr}} < 1\right\}$.

Now consider tracking \eqref{eq: sysmodel} over a noisy channel. Intuitively, the desired anytime exponent is determined only by the growth of the tracking error in the absence of measurements, which by Lemma \ref{lem: cubTimeUpdate} is governed by $\Fbar$. This is independent of whether or not control input is available at the observer. This explains the value of $\beta_{n} = \beta^f_{n} = \frac{2}{n}\log_2\lambda(\Fbar)$ in Theorems \ref{thm: cuboidalThm} and \ref{thm: cuboidalThmWF}. Making this argument rigorous is simple and has not been presented here due to space limitations. 


\section{Simulations}
\label{sec: Simulations}
We present two examples, one scalar and one vector, and stabilize them over a binary erasure channel with erasure probability $\epsilon = 0.3$. The number of channel uses per measurement is fixed to $n=15$. In both cases, time invariant codes $\mathbb{H}_{15,R}\in\mathbb{TZ}_{\frac{1}{2}}$, for an appropriate rate $R$, were randomly generated and decoded using Algorithm \ref{alg: algorithm}.
\subsection{Example 1}
\begin{figure}
\centering
\subfigure[Open loop trajectory]{\includegraphics[scale=0.095]{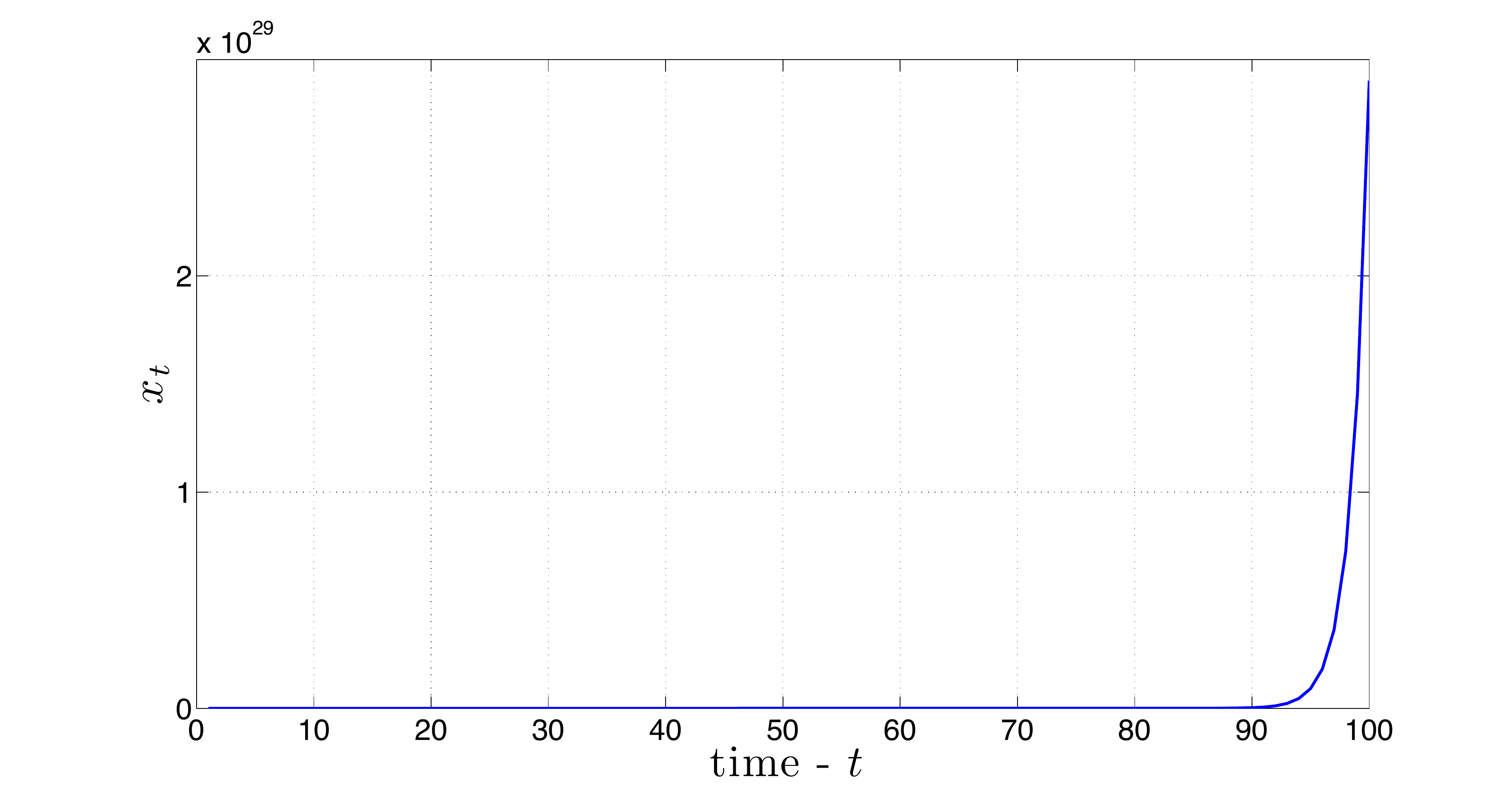}}\,\,
 \subfigure[Trajectory after closing the loop]{\includegraphics[scale=0.095]{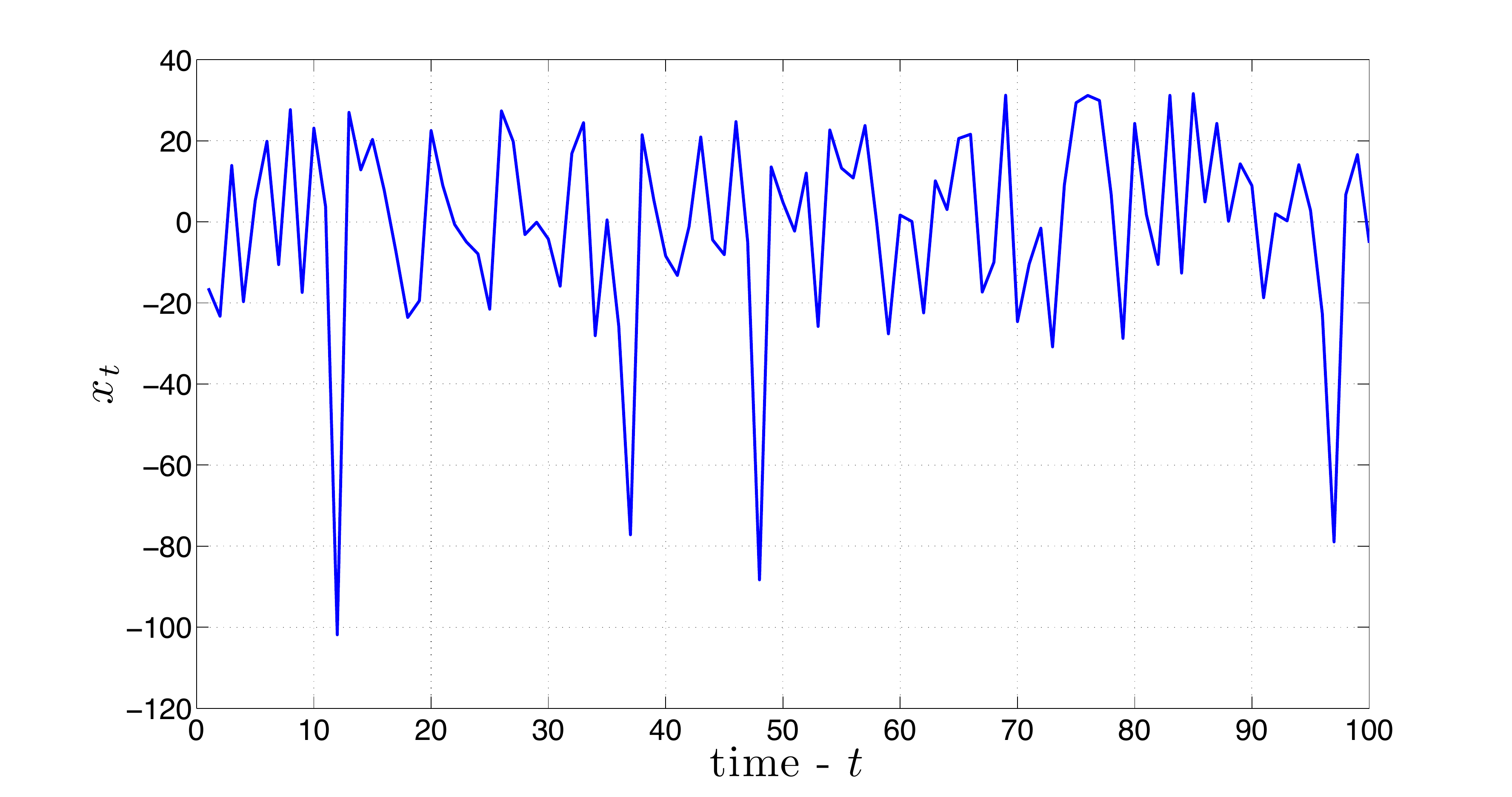}}
\caption{}
\label{fig: samplePath}
\end{figure}
\begin{figure}
 \centering
\includegraphics[keepaspectratio=false,height=4cm,width=7cm]{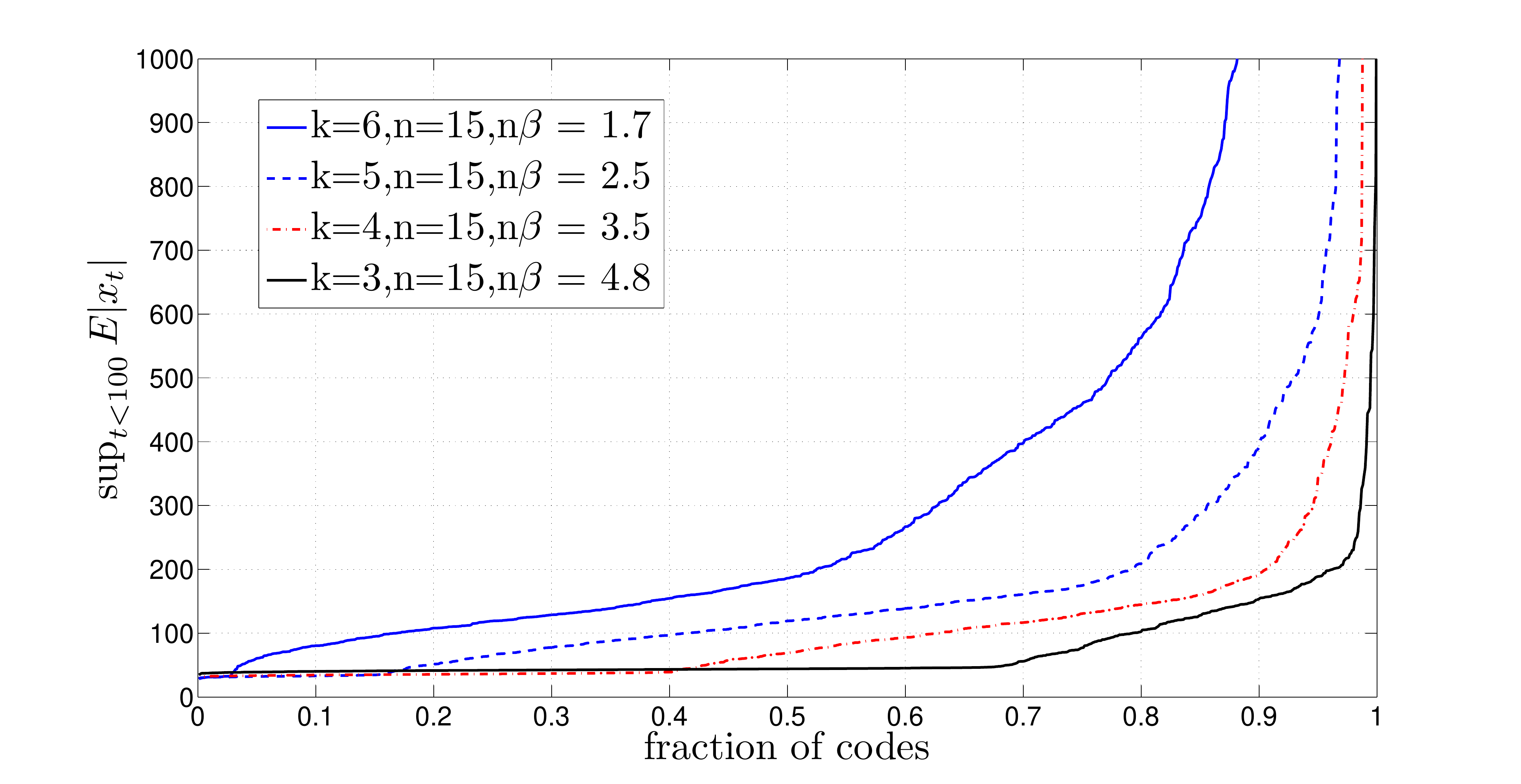}
\caption{The control performance of the code ensemble improves as the rate decreases}
\label{fig: comparison}
\end{figure}

Consider stabilizing the scalar unstable process obtained by setting $m=1$, $-a_1=2$ in \eqref{eq: sysmodel} with $w_t$ and $v_t$ being uniform on $[-30,30]$ and $[-1,1]$ respectively. Using Theorem \ref{thm: cuboidalThm}, inorder to stabilize $x_t$ in the first moment sense, one needs a code with exponent $\beta \geq \frac{1}{n} = 0.0667$ and $k = nR \geq 1$. Using Theorem \ref{thm: Toeplitz}, causal linear codes exist for $\beta <\beta^*= H^{-1}(1-R)\bra{\log_2\bra{\frac{1}{\bh}} + \log_2\bra{2^{1-R}-1}}$. A quick calculation shows that for $k=6,n=15$, $\beta^* = \frac{1.1413}{n} = 0.0761 > 0.0667$. The observer does not have access to the control inputs, so an $s^k-$regular lattice quantizer with bin-width $\delta$ was used to quantize the measurements. The control input is just $u_t = -\hat{x}_{t|t-1}$. The four curves in Fig \ref{fig: comparison} correspond to the following sets of values: ($k=3,\delta=16$), ($k=4,\delta=8$), ($k=5,\delta=4$) and ($k=6,\delta=2$). Fig \ref{fig: samplePath} shows the plot of a sample path of the above process with $k=3,\delta=16,L=2^3$ before and after closing the loop, the fact that the plant has been stabilized is clear.
By easing up on the rate $R$, i.e., by performing coarser quantization but better error correction, the control performance of the code ensemble improves. This is demonstrated in Fig \ref{fig: comparison}. For each value of $k$ from 3 to 6, 1000 time invariant codes were generated at random from $\mathbb{TZ}_{\frac{1}{2}}$. Each such code was used to control the process above over a time horizon of $T=100$. The $x-$axis denotes the proportion of codes for which $\sup_{t<100}\E|x_t|$ is below a prescribed value, e.g., with $k=6,n=15$, $\sup_{t<100}\E|x_t|$ was less than 200 for $50\%$ of the codes while with $k=3,n=15$, this fraction increases to more than $95\%$. The $y-$axis has been capped at 1000 for clarity. This shows that one can tradeoff utilization of communication resources and control performance. 
\subsection{Example 2}
\begin{figure}
 \centering
\includegraphics[keepaspectratio=false,height=4cm,width=8cm]{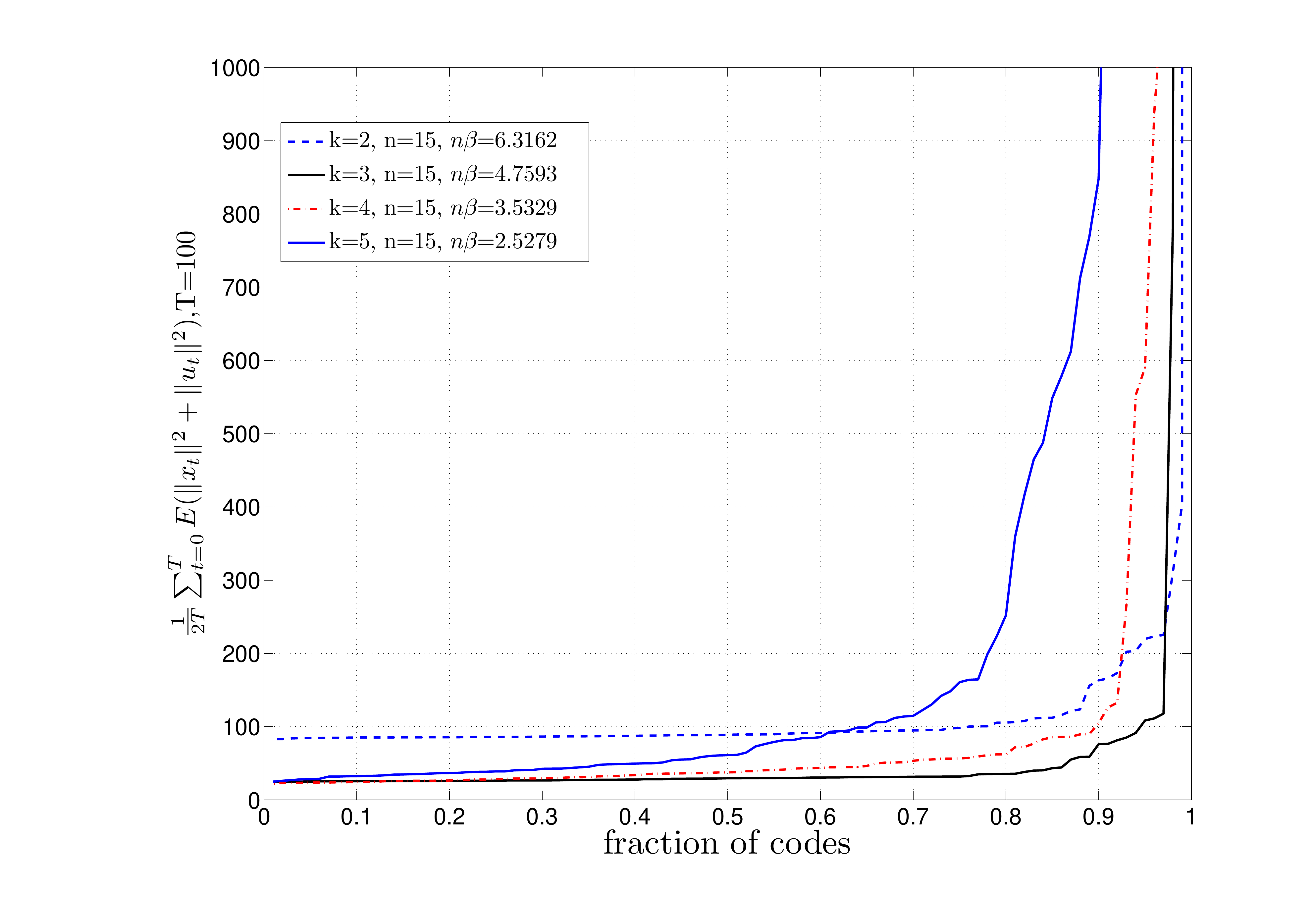}
\caption{The CDF of the LQR costs for different realizations of the codes}
\label{fig: 3DimSys}
\end{figure}

Consider a 3-dimensional unstable system \eqref{eq: sysmodel} with $a_1 = -2$, $a_2 = -0.25$, $a_3 = 0.5$
and $B = \mathcal{I}_3$. Each component of $w_t$ and $v_t$ is generated i.i.d $N(0,1)$ and truncated to [-2.5,2.5]. The eigen values of $F$ are $\{2,-0.5,0.5\}$ while $\lambda(\Fbar) = 2.215$. The observer has access to the control inputs and we use the hypercuboidal filter outlined in Section \ref{sec: proofCuboidal}. Using Theorem \ref{thm: cuboidalThmWF}, the minimum required bits and exponent are given by $k = nR \geq 2$ and $n\beta\geq 2\log_2 2.215 = 2.29$. The control input is $u_t = -\hat{x}_{t|t-1}$. For $k\leq 5$, $n\beta \geq 2.53$. The competition between the rate and the exponent in determining the LQR cost is evident when we look at the LQR cost $\frac{1}{200}\sum_{i=1}^{100}\E\left[\Vert x_t\Vert^2 + \Vert u_t\Vert^2\right]$ in Fig \ref{fig: 3DimSys}. When $k = 2$, the error exponent $n\beta = 6.3$ is large. So, at any time $t$, the decoder decodes all the source bits $\{b_\tau\}_{\tau\leq t-1}$ with a high probability. Hence, the limiting factor on the LQR cost is the resolution the source bits $b_t$ provide on the measurements. But when $k = 5$, the measurements are available almost losslessly but the decoder makes errors in decoding the source bits. Fig \ref{fig: 3DimSys} suggest that the best choice of rate is $R = 3/15 = 0.2$.

\section{Conclusion}
\label{sec: Conclusion}

We presented an explicit construction of anytime reliable tree codes with efficient encoding and decoding over erasure channels. We also gave several sufficient conditions on the rate and reliability required of the tree code to guarantee stability, and argued that they are asymptotically tight. Although the work described here is a major step towards controlling plants over noisy channels, there are many issues to study and resolve. The tradeoff between rate and reliability (how finely to quantize the measurements vs. how much error protection to use) to optimize system performance (such as an LQR cost) remains to be studied, as well as how best to quantize and generate control signals. Furthermore, the problem of constructing efficiently decodable tree codes for other classes of channels, such as the BSC and the AWGNC, remains open.
\bibliographystyle{IEEEbib}
\bibliography{arXivCDC2011}
\def\mIu{\mathcal{I}_u}
\appendix

\subsection{The Minimum Volume Ellipsoid}
\begin{lem}[Theorem 6.1 \cite{Guler}]
\label{lem: minVol}
 The minimum volume ellipsoid $\mE(\hat{P},c)$ covering 
\begin{align*}
\left\{x\in\Re^m| x\in\mE(P,0),\gamma\sqrt{h^TPh}\leq \langle h,x\rangle\leq\delta\sqrt{h^TPh}\right\} 
\end{align*}
where $|\delta| \geq |\gamma|$, is given by
\begin{align}
 \hat{P} = bP - (b-a)\frac{Phh^TP}{h^TPh},\,\,\,c = \xi \frac{Ph}{\sqrt{h^TPh}}
\end{align}
where 
\begin{enumerate}
 \item If $\gamma\delta < -\frac{1}{m}$, then $\xi = 0$, $a = b = 1$
 \item If $\gamma+\delta = 0$ and $\gamma\delta > -\frac{1}{m}$, then
\begin{align*}
 \xi = 0,\,\,a = m\delta^2,\,\,b = \frac{m(1-\delta^2)}{m-1}
\end{align*}
 \item If $\gamma+\delta\neq 0$ and $\gamma\delta > -\frac{1}{m}$, then
\begin{align*}
 \xi &= \frac{m(\gamma+\delta)^2 + 2(1+\gamma\delta) - \sqrt{D}}{2(m+1)(\gamma+\delta)}\\
a &= m(\xi-\gamma)(\xi-\delta),\,\,b = \frac{a-a\gamma^2}{a-(\xi-\gamma)^2}\\
\text{where }D&= m^2(\delta^2 - \gamma^2)^2 + 4(1-\gamma^2)(1-\delta^2)
\end{align*}
\end{enumerate}
\end{lem}
If $|\delta| < |\gamma|$, change $x$ to $-x$ and apply the above result.

\subsection{Upper bounds on $R^f_n$ and $R^f_{e,n}$}

There are several bounds in the Mathematics literature on the roots of a polynomial in terms of the polynomial coefficients, a standard and near optimal bound being the Fujiwara's bound which we state below.
\begin{lem}[Fujiwara's Bound]
 \label{lem: fujiwara}
 Consider the monic polynomial with complex coefficients $f(x) = x^m + c_1x^{m-1}+\ldots+c_m$ and let $\lambda(f)$ denote the largest root in magnitude. Then
\begin{align*}
 \lambda(f) \leq K(f) = 2\max\left\{|c_1|,|c_2|^{\frac{1}{2}},\ldots,|c_{m-1}|^{\frac{1}{m-1}},\left|\frac{c_m}{2}\right|^{\frac{1}{m}}\right\}
\end{align*}
\end{lem}

The upper bounds on $R^f_n$ and $R^f_{e,n}$ can now be proved as follows. The characteristic polynomial of $\Fbar D_{nr}$ is given by $f_{c,nr}(x) = x^m - 2^{-nr}\sum_{i=1}^m|a_i|x^{m-i}$. Applying Lemma \ref{lem: fujiwara}, if the rate $R$ is larger than the smallest value of $r$ that will make $K(f_{c,nr})<1$, then $\lambda(\Fbar D_{nR}) \leq K(f_{c,nR}) < 1$ making $\Fbar D_{nR}$ stable. The bound for $R^f_n$ is then immediate while the bound for $R^f_{e,n}$ follows by noting that the characteristic polynomial of $\Fbar D_{m,nr}$ is $x^m - \sqrt{m}2^{-nr}\sum_{i=1}^m \theta^{i-1}|a_i|x^{m-i}$.

\subsection{The Limiting Case}
\label{subsec: limitingCase}

Let $F$ is any $m$-dimensional square matrix and $f(x)$ denotes its characteristic polynomial. Then the following bounds hold (for details see \cite{Sluis})
\label{eq: sluis}
\begin{align}
\label{eq: sluis1}  \lambda(F) \leq \lambda(\Fbar) \leq \frac{\lambda(F)}{\sqrt[m]{2}-1},\,\,K(f) \leq 2\lambda(\Fbar)
\end{align}
The proof for $\lim_{n\rightarrow\infty}\beta^f_{n} = \beta^*$ follows easily from the first bound in \eqref{eq: sluis1}. 

By the hypothesis of the Lemma, the eigen values of $F_n$ are of the form $\{\mu_i^n\}_{i=1}^m$. To emphasize the fact that $F$ depends on $n$, we write it as $F_n$ and $a_i$ as $a_{i,n}$. Recall that the characteristic polynomial of $F_n$ is given by $f_n(x) = x^m + a_{1,n}x^{m-1}+\ldots+a_{m,n}$. Let $\mIu\triangleq\{i\,\,\arrowvert\,\, |\mu_i| \geq 1\}$, then the following is easy to prove
\begin{align}
\label{eq: appendix1}
\lim_{n\rightarrow\infty}\frac{|a_{i,n}|}{\left|a_{|\mIu|,n}\right|} = 0,\,i\neq |\mIu|,\,\,\lim_{n\rightarrow\infty}\frac{1}{n}\log_2\left|a_{|\mIu|,n}\right| = \sum_{i\in\mIu}\log_2|\mu_i|
\end{align}
We will prove that $R_n$ and $R^f_n$ converge to $R^*$, the proof for $R_{e,n}$ and $R^f_{e,n}$ is similar. From \eqref{eq: appendix1}, it is obvious that $\lim_{n\rightarrow\infty}R_n = \sum_{i\in\mIu}\log_2|\mu_i|$. It remains to show that the limit holds for $R^f_n$. The characteristic polynomial of $\Fbar D_{nr}$ is given by $f_{c,nr}(x) = x^m - 2^{-nr}\sum_{i=1}^m|a_i|x^{m-i}$.
From \eqref{eq: sluis}, we have $\frac{1}{2}K(f_{c,nr}) \leq \lambda(\Fbar D_{nr}) \leq K(f_{c,nr}))$.
Define $R^f_{n,1} \triangleq \argmin_r \left\{\frac{1}{2}K(f_{c,nr}) \leq 1\right\}$ and $R^f_{n,2} \triangleq \argmin_r \left\{K(f_{c,nr}) \leq 1\right\}$.
Then, it is obvious that $ R^f_{n,1}\leq R^f_{n}\leq R^f_{n,2}$. Using Lemma \ref{lem: fujiwara}, some simple algebra and taking limit $n\rightarrow\infty$, we get
\begin{align*}
 \lim_{n\rightarrow\infty}R^f_{n} = \lim_{n\rightarrow\infty}\frac{1}{n}\log_2\max\left\{\frac{|a_{m,n}|}{2},\max_{1\leq i\leq m-1}|a_{i,n}|\right\}
\end{align*}
Combining this with \eqref{eq: appendix1}, we get the desired result, i.e., $\lim_{n\rightarrow\infty}R^f_n = \sum_{i\in\mIu}\log_2|\mu_i|$. 

\end{document}